\documentclass[a4paper]{article}

\usepackage{INTERSPEECH2020}
\usepackage{amsmath}
 \usepackage{booktabs}
 \usepackage{tabu}
 \usepackage{multirow}
 \usepackage{multicol}
 \usepackage{xcolor,colortbl}
 \usepackage{color,soul}
 
\newcolumntype{?}{!{\vrule width 1.1pt}}
\newcommand{\hbl}{\noalign{
\hrule height 1.1pt
}}
\newcommand{\pad}{1.3}
\title{Learning not to Discriminate: Task Agnostic Learning for Improving Monolingual and Code-switched Speech Recognition}

\name{Gurunath Reddy M\textsuperscript{$+$}, Sanket Shah\textsuperscript{$+$}, Basil Abraham\textsuperscript{$*$},\\
Vikas Joshi\textsuperscript{$*$}, Sunayana Sitaram\textsuperscript{$+$}}
\address{\textsuperscript{$+$}Microsoft Research India,\\
\textsuperscript{$*$}Microsoft Corporation}
\email{\{t-gumadh, t-sansha, basil.abraham, vikas.joshi, sunayana.sitaram\}@microsoft.com}

\begin{document}

\maketitle
\begin{abstract}

Recognizing code-switched speech is challenging for Automatic Speech Recognition (ASR) for a variety of reasons, including the lack of code-switched training data. Recently, we showed that monolingual ASR systems fine-tuned on code-switched data deteriorate in performance on monolingual speech recognition, which is not desirable as ASR systems deployed in multilingual scenarios should recognize both monolingual and code-switched speech with high accuracy. Our experiments indicated that this loss in performance could be mitigated by using certain strategies for fine-tuning and regularization, leading to improvements in both monolingual and code-switched ASR. In this work, we present further improvements over our previous work by using domain adversarial learning to train task agnostic models. We evaluate the classification accuracy of an adversarial discriminator and show that it can learn shared layer parameters that are task agnostic. We train end-to-end ASR systems starting with a pooled model that uses monolingual and code-switched data along with the adversarial discriminator. Our proposed technique leads to reductions in Word Error Rates (WER) in monolingual and code-switched test sets across three language pairs.

\end{abstract}
\noindent\textbf{Index Terms}: speech recognition, code-switching, adversarial learning, transfer learning

\section{Introduction}

Recognizing code-switched speech is challenging for Automatic Speech Recognition (ASR) systems due to the lack of large amounts of labeled code-switched speech and text data for training Acoustic and Language Models. Recently, we showed that even if there is sufficient code-switched speech data to train models, there is a loss in performance on monolingual test sets when monolingual models are trained or fine-tuned with code-switched data \cite{shah2020learning}. Since code-switched and monolingual speech co-occur, it is imperative that models perform well on code-switched speech while not deteriorating on monolingual speech.

With this goal in mind, in \cite{shah2020learning} we proposed strategies for learning how to recognize code-switched speech while not forgetting monolingual speech recognition in the following scenarios:\\ \textbf{Case 1}: If monolingual and code-switched data are both available and a model can be retrained from scratch, regularization strategies and fine-tuning a pooled model that uses all data leads to best results across data sets.\\
\textbf{Case 2}: If only a monolingual model is available and a new model cannot be retrained from scratch, the Learning Without Forgetting \cite{Li2018LearningWF} framework can be used to improve performance on all test sets compared to a monolingual model fine-tuned on code-switched data.

In this work, we build upon our findings for Case 1, in which we have access to both monolingual and code-switched data and can retrain a model from scratch. When we train a joint model to learn both monolingual and code-switched speech recognition tasks with task specific and shared layer parameters, the model tends to drift towards one particular task. This drift is because shared layers try to learn task specific features which is not ideal for a joint model that needs to perform well on both tasks. Hence, we need to learn task invariant or agnostic shared layer parameters which lead to task agnostic features at shared layers and discriminant features at task specific layers.

In this work, we learn task agnostic shared layer parameters by adversarial discriminative learning. We show that it is possible to improve performance by using adversarial learning over our previously proposed techniques of fine-tuning and regularization on monolingual and code-switched test sets that span three language pairs - Tamil-English, Telugu-English and Gujarati-English. In this work, we assume that there exists a classifier that will classify code-switched and monolingual utterances prior to recognition by our model, however, our technique can also be used if this assumption does not hold.

The rest of the paper is organized as follows. Section 2 relates our work to prior work. Section 3 describes our experimental setup and results. Section 4 concludes.

\section{Relation to Prior Work}

In this paper, we learn task agnostic shared layer parameters by adversarial learning inspired by the domain-adversarial training of neural networks~\cite{ganin2016domain}. Originally, domain-adversarial learning was proposed to adapt models trained on labeled data to new unlabeled data by adversarial discrimination. Adversarial training has been adopted recently for many tasks:~\cite{ganin2016domain} and~\cite{tzeng2017adversarial} use  adversarial learning for domain adaptation for image classification. ~\cite{chen2017adversarial} utilized adversarial strategies for word boundary segmentation of the Chinese heterogeneous data.~\cite{shinohara2016adversarial} and~\cite{saon2017english} utilized adversarial learning for environment and speaker adaptation for robust speech recognition. Recently,~\cite{yi2018language} explored adversarial learning for transferring knowledge from source language to target language for low-resource ASR models.

The following approaches have been explored for end-to-end code-switched speech recognition. A hybrid attention based architecture is described in~\cite{Luo2018TowardsEC} for Mandrian-English code-switched ASR. Multiple fine-tuning approaches have been studied to improve code-switched speech recognition in~\cite{Winata2020MetaTransferLF} and~\cite{choudhury-etal-2017-curriculum}.  Multi-task learning strategies have also been proposed for improving code-switched speech recognition in~\cite{cs-multi-task} and~\cite{Song2017AML}. Recently, we proposed approaches to learn code-switched speech recognition without forgetting monolingual speech recognition \cite{shah2020learning} using various regularization and fine-tuning strategies, as well as the Learning Without Forgetting \cite{Li2018LearningWF} framework.

\section{Experimental Setup}

\begin{figure*}[!htb]
    \center{\includegraphics[scale=0.75]{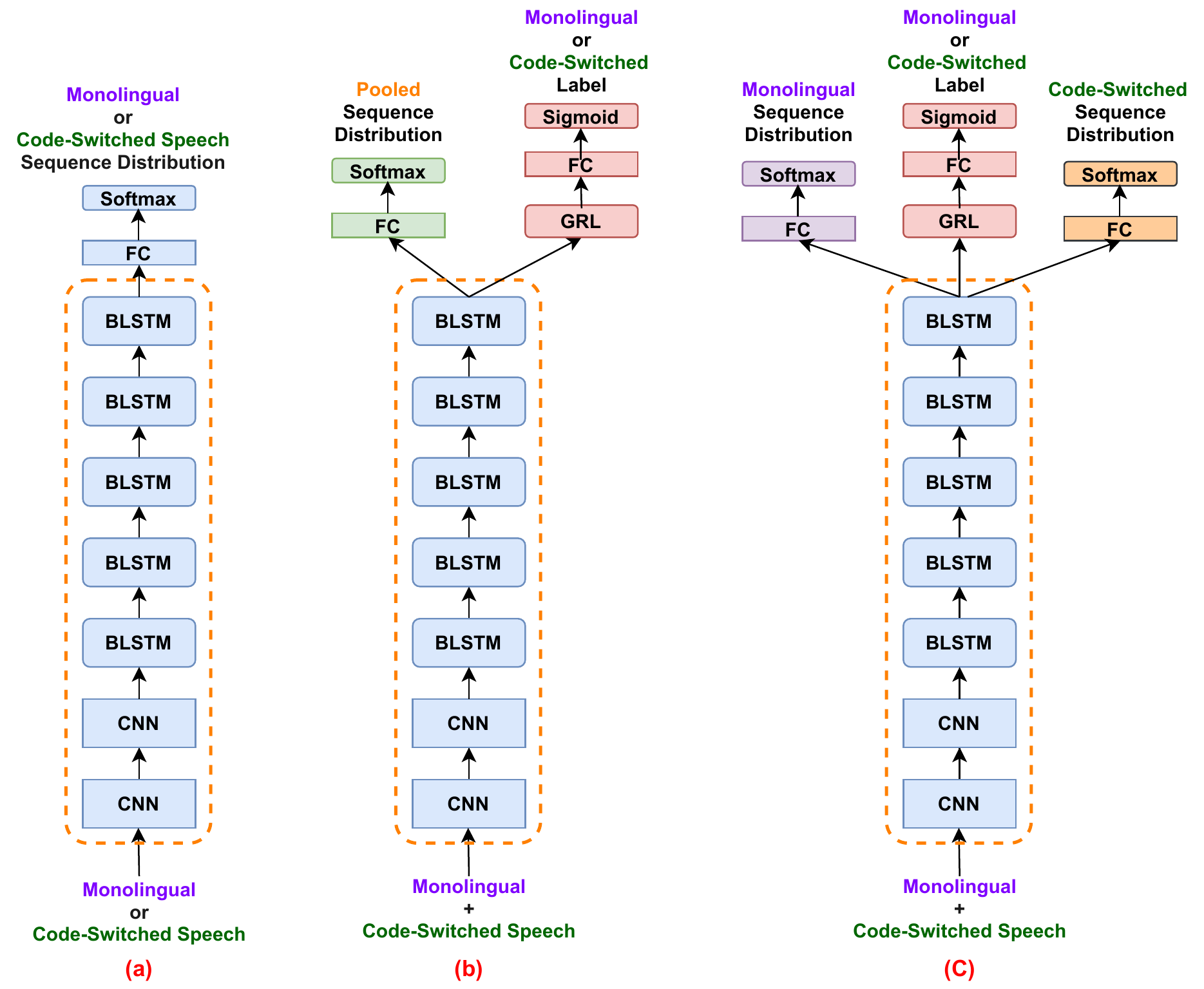}}
    \caption{Illustration of the proposed adversarial learning of task agnostic shared layer parameters for monolingual and code-switched speech recognition. (a) Baseline model trained with either monolingual or code-switched speech utterances. (b) Pooled model trained with both monolingual and code-switched speech utterances along with adversarial task (monolingual or code-switched) discriminator to learn task agnostic shared layer parameters. (c) Multi-task adversarial discriminator trained to recognize monolingual and code-switched speech recognition independently. CNN and BLSTM are the shared layers shown inside the dotted rectangular box. Set of FC and Softmax are the task specific layers. GRL, FC and Sigmoid form the adversarial task discriminator layers.     
    \label{fig:adv-diagram}}
\end{figure*}

\subsection{Data}

We use the same data and baselines as described in \cite{shah2020learning}, which we mention in brief in this section. We carried out experiments for three languages - Tamil (TA), Telugu (TE) and Gujarati (GU) and their code-switched counterparts with English - Tamil-English, Telugu-English and Gujarati-English. Although all three languages were mixed with English, the type and extent of mixing was different. We used two types of speech data for training - conversational data as well as phrasal data, which is similar to read speech, while for testing, only phrasal data was used. We test our models on monolingual and code-switched data sets separately to ensure that models perform well on both. Hence, we have six test sets that we evaluate our models on. Table \ref{tab:traintestdata} describes the dataset size in hours.

\begin{table}[!ht]
\caption{Training and test data statistics}
\label{tab:traintestdata}
\centering
\renewcommand{\arraystretch}{\pad}
\resizebox{0.45\textwidth}{!}{
\begin{tabular}{?>{\columncolor[HTML]{EFEFEF}}l|l|l|l|l?}
\hbl
\textbf{} & \cellcolor[HTML]{EFEFEF}\textit{\textbf{Train + Dev}} & \cellcolor[HTML]{EFEFEF}\textit{\textbf{Train + Dev }}& \cellcolor[HTML]{EFEFEF}\textit{\textbf{Test}} &\cellcolor[HTML]{EFEFEF}\textit{\textbf{Test}} \\ 

\textbf{} & \cellcolor[HTML]{EFEFEF}\textit{\textbf{(MONO)}} & \cellcolor[HTML]{EFEFEF}\textit{\textbf{(CS) }}& \cellcolor[HTML]{EFEFEF}\textit{\textbf{(MONO) }} &\cellcolor[HTML]{EFEFEF}\textit{\textbf{(CS) }} \\ \hline
TA& 212 hrs & 177 hrs (CMI: 22.08) & 24 hrs  & 19 hrs (CMI: 17.07)\\ \hline
TE& 170 hrs    & 243 hrs (CMI: 23.85) & 19 hrs  & 28 hrs (CMI: 21.62)                  \\ \hline
GU& 241 hrs& 186 hrs (CMI: 18.91)& 26 hrs  & 18 hrs (CMI: 16.32)             \\ 
\hbl
\end{tabular}
}
 \end{table}

The Code Mixing Index (CMI) \cite{gamback2014measuring} measures the amount of code-switching in a corpus by using word frequencies. We measure the CMI of our code-switched train and test sets and report them in parentheses in Table \ref{tab:traintestdata}. The CMI of Telugu-English is the highest, while Gujarati-English is the lowest suggesting that Telugu-English is the most code-switched while Gujarati-English is the least code-switched among the languages under consideration.




\begin{table}[!ht]
\caption{Baseline Word Error Rates (WER in \%)}
\label{tab:baselines123}
\renewcommand{\arraystretch}{\pad}
\centering
\begin{tabular}{?>{\columncolor[HTML]{EFEFEF}}l|l|l|l?}
\hbl
\cellcolor[HTML]{EFEFEF}\textbf{Test Set}        & \cellcolor[HTML]{EFEFEF}\textit{\textbf{Exp1}} & \cellcolor[HTML]{EFEFEF}\textit{\textbf{Exp2}} & \cellcolor[HTML]{EFEFEF}\textit{\textbf{Exp3}} \\ \hline
TA-MONO & 50.09 & 70.20 & \textbf{48.47}\\\hline
TA-CS   & 67.62 & 63.70 & \textbf{55.93}\\ \hline
TE-MONO & 46.90 & 57.52 & \textbf{45.15} \\\hline
TE-CS   & 59.91 & 44.46 & \textbf{40.75}\\ \hline
GU-MONO & 41.99 & 54.83 & \textbf{40.96}\\ \hline
GU-CS   & 51.68 & 47.50 & \textbf{45.92}\\ 
\hbl
\end{tabular}
\end{table}

\subsection{Baseline experiments}

We denote our training monolingual datasets {($X_1^M$, $Y_1^M$),...,($X_n^M$, $Y_n^M$)} where $M$ $\in$ \{TE/TA/GU\}, code-switched datasets {($X_1^{CS}$, $Y_1^{CS}$),...,($X_n^{CS}$, $Y_n^{CS}$)} where $CS$ $\in$ \{TE-EN/TA-EN/GU-EN\}. The labels $Y$ are graphemes and the character set includes the union of English and the respective language's characters. Further, we denote $T = \{M, CS\}$ where $T$ is the monolingual or code-switched speech recognition task.    

Our baseline model consists of two Convolution Neural Network (CNN) layers followed by five bidirectional long-short term (BLSTM) layers of 1024 dimension. These parameters are shared between the monolingual and code-switched task and are denoted by $\theta_s$. Further, the frame-wise posterior distribution is conditioned on the input frame $X^T_i$, is calculated by a forward pass through the shared layers, $\theta_s$ and through a fully-connected layer, $\theta_T$ followed by softmax computation over labels as shown in Fig.~\ref{fig:adv-diagram}(a). We maximize the conditional posterior distribution by minimizing the Connectionist Temporal Classification (CTC) \cite{10.1145/1143844.1143891} criterion represented by $L_T(\theta_s, \theta_T)$.      


\noindent The model parameters are trained using stochastic gradient descent (SGD) optimizer. The learning rate ($\lambda$) is initialized with 3e-4. The model is trained for 40 epochs, with mini-batch size equal to 64 per GPU. The model parameters are updated using the back propagation algorithm.




We evaluated our proposed approach against three baselines, which we refer to as Exp1, Exp2 and Exp3.

\begin{itemize}
    \item \textbf{Exp1}: Monolingual-only baseline, consisting of models trained only on monolingual data 
    \item \textbf{Exp2}: Code-switched-only baseline, consisting of models trained only on code-switched data
    \item \textbf{Exp3}: Pooled model, consisting of models trained using all the data from Exp1 and Exp2
\end{itemize}

An n-gram Language Model (LM), trained using transcriptions from the training data is used during decoding. Table \ref{tab:baselines123} shows Word Error Rates (WER) of all three baselines on both monolingual and code-switched test sets. Exp3, which is the pooled model consisting of monolingual and code-switched data performs best on all test sets. Exp1 performs better on monolingual test sets than Exp2, and the reverse is true for code-switched test sets, as expected. 

\subsection{Adversarial task agnostic pooled model}

From the baseline experiments we observe that the pooled model performs better than the monolingual or code-switched model for all test sets. Even though the pooled model performs significantly better than code-switched only baseline, the improvements are only marginally better on monolingual test sets compared to the monolingual only baseline. 

We hypothesize that this is because shared layer parameters learn unwanted task specific features which drifts the performance of the monolingual recognition task towards the code-switched task. In order to alleviate this performance drift, we propose learning task invariant shared layer parameters in the pooled model by adversarial training as shown in Fig.~\ref{fig:adv-diagram}(b). The adversarial pooled model consists of task independent shared ($\theta_s$) layers, task dependent ($\theta_T$) layers and adversarial task discriminator consisting of a fully connected (FC) layer, gradient reversal layer (GRL) and sigmoid activation. The parameters of the adversarial discriminator are denoted by $\theta_a$.

The gradient reversal layer of the adversarial task discriminator ensures that features from the shared layers are as in-discriminant as possible for the given task so that the shared layers learn a generalized representation. The GRL contains no trainable parameters and acts as a identity transformer during forward pass. However, during back-propagation the GRL reverses the gradients of the previous layers i.e., multiplies the gradients by -1 and passes it to the next layers which helps in making the shared layer features in-discriminant to specific task. For each utterance $u$, the adversarial task discriminator is trained to discriminate speech utterances into either monolingual or code-switched ($T = \{M, CS\}$).

\begin{equation}
L_A(\theta_s, \theta_a) = - \sum_{u=1}^{N} \log P (T_u | X_u^T; \theta_s, \theta_a) 
\end{equation}

\noindent where $X_u^T$ and $T_u$ represents $u_{th}$ input utterance and corresponding label and $N$ represents the total numbers of utterances in the dataset. Even though the discriminator is trained to minimize the classification loss, the gradients of the discriminator is negative so that the shared layers are trained to be task independent. The parameters of the adversarial task discriminator are updated as 
\begin{equation}
\theta_s \longleftarrow \theta_s + \lambda \frac{\partial L_A}{\partial \theta_s}   
\end{equation}

\begin{equation}
\theta_a \longleftarrow \theta_a - \lambda \frac{\partial L_A}{\partial \theta_a}   
\end{equation}

\noindent In our previous work \cite{shah2020learning}, we found that speech recognition performance can be improved by initializing the parameters of the model from a pretrained model. Hence, shared layer parameters of the adversarial task agnostic pooled model is initialized from the baseline pooled model (Exp3) and trained with the loss function

\begin{equation}
L_{AP}(\theta_s, \theta_T, \theta_a) = L_T(\theta_s, \theta_T) + L_A(\theta_s, \theta_a) 
\end{equation}

\noindent The parameters of the model are updated as 

\begin{equation}
\theta_s \longleftarrow \theta_s - \lambda \left( \frac{\partial L_T}{\partial \theta_s} - \frac{\partial L_A}{\partial \theta_s} \right)
\end{equation}

\begin{equation}
\theta_T \longleftarrow \theta_T - \lambda \frac{\partial L_T}{\partial \theta_T}   
\end{equation}

\begin{equation}
\theta_a \longleftarrow \theta_a - \lambda \frac{\partial L_A}{\partial \theta_a}   
\end{equation}

The performance of the adversarial task agnostic pooled model (Exp5) compared with the pooled model (Exp3) is shown in Table~\ref{tab:pooled_adversarial}. We can see that Exp5 performs better than Exp3 for all test sets. This indicates that the model benefits from adversarial task-discriminative training for improving  both monolingual and code-switched speech recognition.

\subsection{Multi-task adversarial speech recognition model}

In our previous work \cite{shah2020learning}, we observed that a multi-task model trained with separate monolingual and code-switched task specific layers yields better performance than the pooled model. Hence, we trained a joint monolingual and code-switched multi-task model as shown in Fig.~\ref{fig:adv-diagram}(c). The multi-task adversarial speech recognition model consists of shared layers ($\theta_s$), task specific monolingual ($\theta_m$) and code-switched ($\theta_c$) layers, and the adversarial task discriminator ($\theta_a$) as shown in Fig~\ref{fig:adv-diagram}(c). The shared layer parameters of the multi-task model are initialized from the pooled model as before and trained jointly end-to-end along with the adversarial task discriminator.

\begin{equation}
L_{MA}(\theta_s, \theta_m, \theta_c, \theta_a) = L_M(\theta_s, \theta_m) + L_{CS}(\theta_s, \theta_c) + L_A(\theta_s, \theta_a) 
\end{equation}

\noindent where $L_{M}(\theta_s, \theta_m)$, and $L_{CS}(\theta_s, \theta_c)$ are the individual monolingual and code-switched loss functions. Similar to Exp5, the utterance level adversarial loss $L_A(\theta_s, \theta_a)$ makes the shared layer features as in-discriminant as possible to monolingual and code-switched speech utterances while learning discriminant features at task specific monolingual and code-switched private layers by updating the parameters

\begin{equation}
\theta_s \longleftarrow \theta_s - \lambda \left( \frac{\partial L_M}{\partial \theta_s} - \frac{\partial L_{CS}}{\partial \theta_s} - \frac{\partial L_A}{\partial \theta_s} \right)
\end{equation}

\begin{equation}
\theta_m \longleftarrow \theta_m - \lambda \frac{\partial L_M}{\partial \theta_m}   
\end{equation}

\begin{equation}
\theta_c \longleftarrow \theta_c - \lambda \frac{\partial L_{CS}}{\partial \theta_c}   
\end{equation}

\begin{equation}
\theta_a \longleftarrow \theta_a - \lambda \frac{\partial L_A}{\partial \theta_a}   
\end{equation}

The performance comparison of Multi-task adversarial model (Exp6), adversarial task agnostic pooled model (Exp5), the best fine-tuned pooled model ~\cite{shah2020learning} (Exp4), and the pooled model (Exp3) are shown in Table~\ref{tab:pooled_adversarial}. We can see that the Multi-task adversarial model (Exp6) outperforms all other models on all monolingual and code-switched test sets. The improved WER can be attributed to the fact that adversarial training helps the shared layer parameters to learn task invariant monolingual and code-switched features, and having task specific layers further help in improving accuracy for individual tasks. 

An important caveat to note here is that task specific layers require knowledge of whether an utterance is code-switched or not. This can be achieved either by using a classifier to classify an utterance as monolingual or code-switched, or the output from both task-specific layers can be averaged to make the final prediction. In future work, we plan to compare the results of both these techniques to the proposed model that uses ground-truth knowledge of monolingual and code-switched utterances. 

\begin{table}[t]
\caption{WER$[\%]$ of pooled (Exp3), fine-tuned pooled model \cite{shah2020learning} (Exp4), adversarial task agnostic pooled model (Exp5) and multi-task adversarial speech recognition model (Exp6)}
\label{tab:pooled_adversarial}
\renewcommand{\arraystretch}{\pad}
\centering
\resizebox{0.35\textwidth}{!}{
\begin{tabular}{?>{\columncolor[HTML]{EFEFEF}}l|l|l|l|l?}
\hbl
\cellcolor[HTML]{EFEFEF}\textbf{Test Set}        & \cellcolor[HTML]{EFEFEF}\textit{\textbf{Exp3}} &
\cellcolor[HTML]{EFEFEF}\textit{\textbf{Exp4}} & \cellcolor[HTML]{EFEFEF}\textit{\textbf{Exp5}} & \cellcolor[HTML]{EFEFEF}\textit{\textbf{Exp6}}  \\  \hline

TA-MONO &48.47 & 48.38 & 47.26& \textbf{46.34}\\\hline
TA-CS   & 55.93 & 54.63 & 54.44& \textbf{53.15} \\\hline
TE-MONO & 45.15 & 44.18 &44.22& \textbf{43.07} \\\hline
TE-CS   & 40.75 &39.32 & 40.19& \textbf{39.19}\\ \hline
GU-MONO & 40.96 & 38.21& 40.11& \textbf{38.20}\\ \hline
GU-CS   & 45.92 & 42.07& 45.56& \textbf{41.53}\\ \hbl
\end{tabular}
}
\end{table}

\subsection{Classification experiments}

In order to test whether the shared layers indeed learn task-invariant parameters, we perform classification experiments. We trained a model with shared layers ($\theta_s$) and speech utterance classifier layers which classifies the utterances into monolingual or code-switched with GRL (adversarial discriminator) and without GRL (vanilla classifier) to observe the effect of GRL on the shared layers. 

\begin{figure}[!htb]
    \center{\includegraphics[scale=0.5]{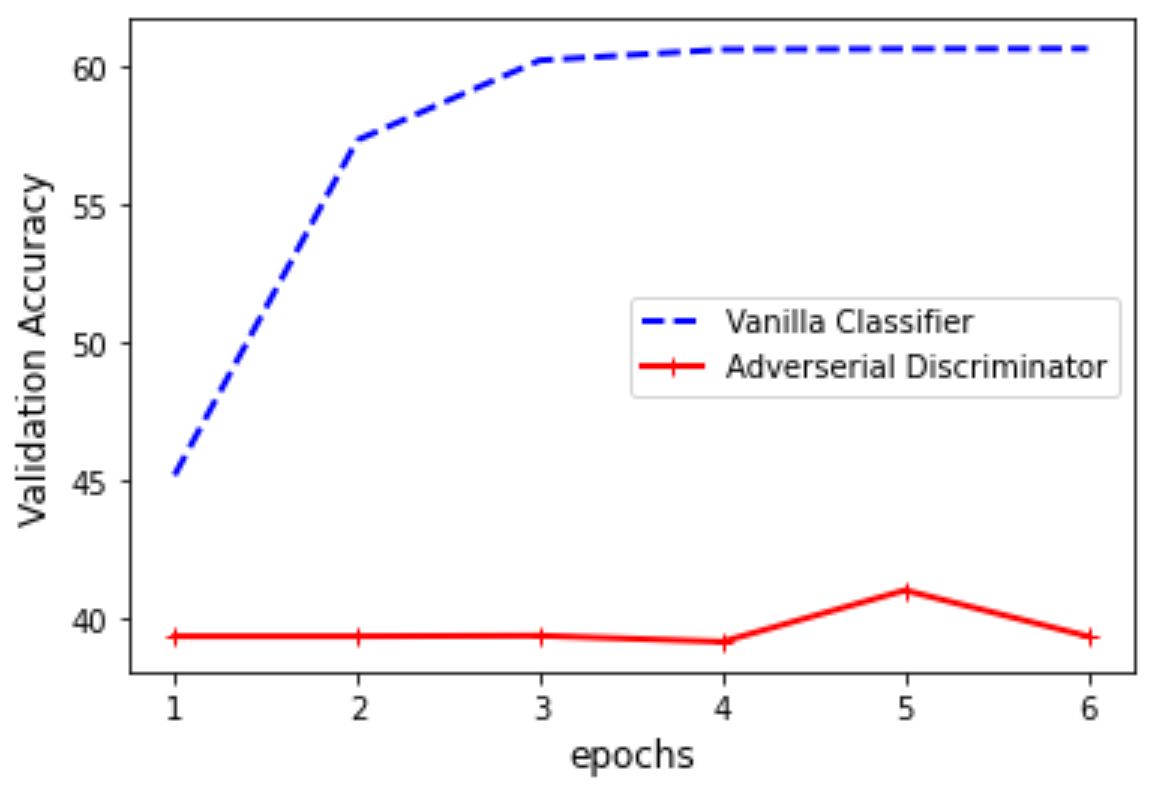}}
    \caption{Validation accuracy of the vanilla classifier and the adversarial discriminator.     
    \label{fig:adv-dis-clas}}
\end{figure}

The validation accuracy of both models is shown in Fig.~\ref{fig:adv-dis-clas}. We observe that the validation accuracy remains constant for the adversarial discriminator, while it keeps increasing for the vanilla classifier. This indicates that the shared layers below the classification layers learn task agnostic features in case of the adversarial discriminator. In contrast, the shared layers learn discriminative features for the vanilla classifier. 

\section{Summary and Conclusions}

Although monolingual and code-switched speech recognition tasks are similar, we see that trying to improve performance on one hampers the performance on the other. Specifically, training a single model with pooled data containing both monolingual as well as code-switched speech performs better than individual models trained on task-specific data. However, gains on monolingual speech recognition are much lower compared to code-switched speech recognition due to the fact that shared layers learn some task-specific features. 

In this paper, we show that learning task invariant shared layer parameters in a pooled model using adversarial training outperforms a pooled model on both monolingual as well as code-switched test sets across three language pairs. We further experiment with adding task specific layers to this model to allow the model to learn some task specific parameters and show improvements on all test sets. Thus, we show that to improve performance on both monolingual as well code-switched speech recognition task, having task invariant shared layers as well as task specific layers are necessary. In future work, we plan to explore techniques to incorporate a Language Identification (LID) system to classify utterances into code-switched or monolingual, as well as explore techniques to use outputs from task specific output layers in the absence of an LID system.

\bibliographystyle{IEEEtran}

\bibliography{mybib}


\end{document}